\documentclass[9pt,twocolumn,twoside]{opticajnl}

\journal{opticajournal} 

\setboolean{shortarticle}{true}

\usepackage{lineno}

\usepackage{graphicx}
\usepackage{subcaption}
\usepackage[normalem]{ulem}
\usepackage{textcomp}
\usepackage{gensymb}
\usepackage{bm}

\title{Goos–H\"anchen Shifts at 458 nm from DUV Dichroic Laser Mirrors}

\author[1]{Shamaila Manzoor}
\author[1,2]{Mauro Chiarotti}
\author[1]{Paul J. Robert}
\author[1,2,,*]{Nicola Poli}

\affil[1]{Dipartimento di Fisica e Astronomia, Università degli Studi di Firenze, Via G. Sansone 1, 50019 - Sesto
Fiorentino, Italy}
\affil[2]{Laboratorio Europeo di Spettroscopia Non-Lineare, LENS Via N. Carrara 1, I-50019 Sesto Fiorentino,
Italy}

\affil[*]{nicola.poli@unifi.it}

\begin{abstract}

We report polarization- and angle-dependent lateral beam shifts observed at 458~nm upon reflection from a deep-ultraviolet (DUV) dichroic multilayer mirror. These  Goos--H\"anchen (GH) shifts originate from the phase of the complex Fresnel reflection coefficient and are quantitatively reproduced using a transfer-matrix model. Simultaneous reflectance (R), transmittance (T), and GH shift measurements provide complementary constraints on the coating's amplitude and phase response. The analysis is extended to transmission, predicting analogous phase-induced lateral shifts. In frequency-doubling cavities using dichroic mirrors, these phase effects can accumulate over multiple reflections, leading to deviations from ray-optics predictions that may affect cavity mode structure and resonance conditions. Our results show a GH shift of as much as 2~\textmu m (\bm{$\approx 4.4\lambda$}), which, combined with RT data, enables improved characterization of the phase response of the multilayer coating. The p-polarized reflectance is minimized at a pseudo-Brewster angle of \bm{$\theta_B \approx 46^\circ$} at 458~nm, providing a natural reference point around which the GH shift is enhanced.
\end{abstract}

\setboolean{displaycopyright}{false} 

\begin{document}
\maketitle
\section{Introduction}

The Goos--H\"anchen (GH) shift is a lateral displacement of an optical beam upon reflection at an interface, arising from the angular dependence of the phase of the complex reflection coefficient. First observed experimentally by Goos and H\"anchen in 1947 for transverse electric (s) polarization~\cite{GHShift1947}, the effect was later extended to transverse magnetic (p) polarization~\cite{GHShift1949}. The effect was theoretically described by Artmann using the stationary-phase approximation which related the shift to the angular derivative of the reflection phase.~\cite{Artmann1948}. Beyond total internal reflection, GH shifts have been investigated in partial reflection and transmission regimes, and are referred to as net GH shift or lateral shift.

Significant effort has been devoted to enhancing and controlling GH shifts through engineered optical phase responses in structured and resonant systems, including photonic crystals~\cite{Soboleva2012}, metasurfaces~\cite{Kong2019}, graphene-based systems~\cite{Zeng2017}, and absorbing media~\cite{Fan2003}. Related beam-shift effects have also been observed in atomic and hybrid atom-optomechanical systems, including Bose--Einstein-condensate-based platforms, where quantum coherence, detuning, and interference provide additional mechanisms to manipulate reflected and transmitted beam shifts~\cite{Abbas2024,din2025}. In all cases, the underlying mechanism is the angular dependence of the complex reflection or transmission phase. Weak measurement techniques~\cite{Jayaswal2013,Santana2016,Santana2019} have further enabled amplification of otherwise small beam shifts through interference between slightly displaced optical beams. Beyond fundamental interest, GH shifts can influence precision optical systems by affecting beam propagation and mode behavior in optical waveguides~\cite{Stenmark2016} and optical microcavities~\cite{Unterhinninghofen2008}. In optical implementations, these effects are relevant in dielectric multilayer coatings, where thin-film interference generates strong angular and polarization dependencies in reflectance, transmittance, and phase. Although GH shifts in idealized multilayer systems have been studied theoretically, experimental investigation in realistic coating systems under partial reflection and transmission conditions relevant to cavity optics remains limited. Bridging this gap is crucial for dichroic mirrors in frequency-conversion and deep-ultraviolet systems, where coatings simultaneously separate harmonic wavelengths and define beam propagation paths. Here, small phase gradients translate into measurable beam shifts missed by ray-optic models, affecting alignment sensitivity and mode propagation. In this work, we experimentally investigate polarization- and angle-dependent GH shifts in a dielectric multilayer mirror at 458 nm under partial reflection. By combining beam-shift measurements with angular reflectance and transmittance data, we constrain the phase response of the complex reflection coefficient and extend the analysis to transmission-induced beam shifts for application-relevant coatings, as for example, those used in frequency-doubling cavities.

\section{Theoretical Background}
\subsection{Beam Propagation Model}
When a finite-width optical beam is reflected from a multilayer dielectric interface, it acquires a small lateral shift relative to geometric optics. Such shift stems from the angular dependence of both the amplitude and the phase of the complex reflection coefficient.

Following a full angular spectrum (Fourier) approach~\cite{goodman1969}, a monochromatic incident beam is decomposed into plane-wave components, characterized by transverse wavevector $k_x$, each reflected complex coefficient $r(k_x)$, obtained from the transfer-matrix method~\cite{BornWolf,yeh2004,macleod2010}. The incident beam shape is modeled as a monochromatic Gaussian beam of waist $w_0$ incident at an angle $\theta$. At the interface plane (z\,=\,0), the transverse electric field is given by~\cite{BornWolf}:

\begin{equation}
E_\mathrm{in}(x) =
\exp\left(-\frac{x^2}{w_0^2}\right)
\exp\left(i k_0 x \sin\theta \right),
\end{equation}

where $k_0 = 2\pi/\lambda$. For oblique incidence, the beam profile on the interface is effectively stretched because of geometric projection. This effect is thus incorporated directly into the numerical representation of the incident field. The beam is decomposed into its angular spectrum via a Fourier transform:

\begin{equation}
E_\mathrm{in}(k_x) =
\frac{1}{\sqrt{2\pi}}
\int_{-\infty}^{\infty}
E_\mathrm{in}(x)\, e^{-i k_x x}\, dx.
\end{equation}

This spectrum is centered at $k_x = k_0 \sin\theta$, with a width determined by the waist of the beam $w_0$. After reflection, each component propagates to a distance $z$ in free space, acquiring a longitudinal phase factor. The reflected field is, therefore:

\begin{equation}
E_\mathrm{ref}(x, z) =
\frac{1}{\sqrt{2\pi}}
\int_{-\infty}^{\infty}
E_\mathrm{in}(k_x)\, r(k_x)\,
e^{i k_x x}\, e^{i k_z z}\, dk_x,
\end{equation}

where $k_z = \sqrt{k_0^2 - k_x^2}$, showing a distortion on the beam profile. Evaluating the propagated reflected field naturally includes both spatial and angular contributions to the observed shift.

\section{Experimental Setup}

\begin{figure}[h!]
\centering
\includegraphics[width=0.8\linewidth]{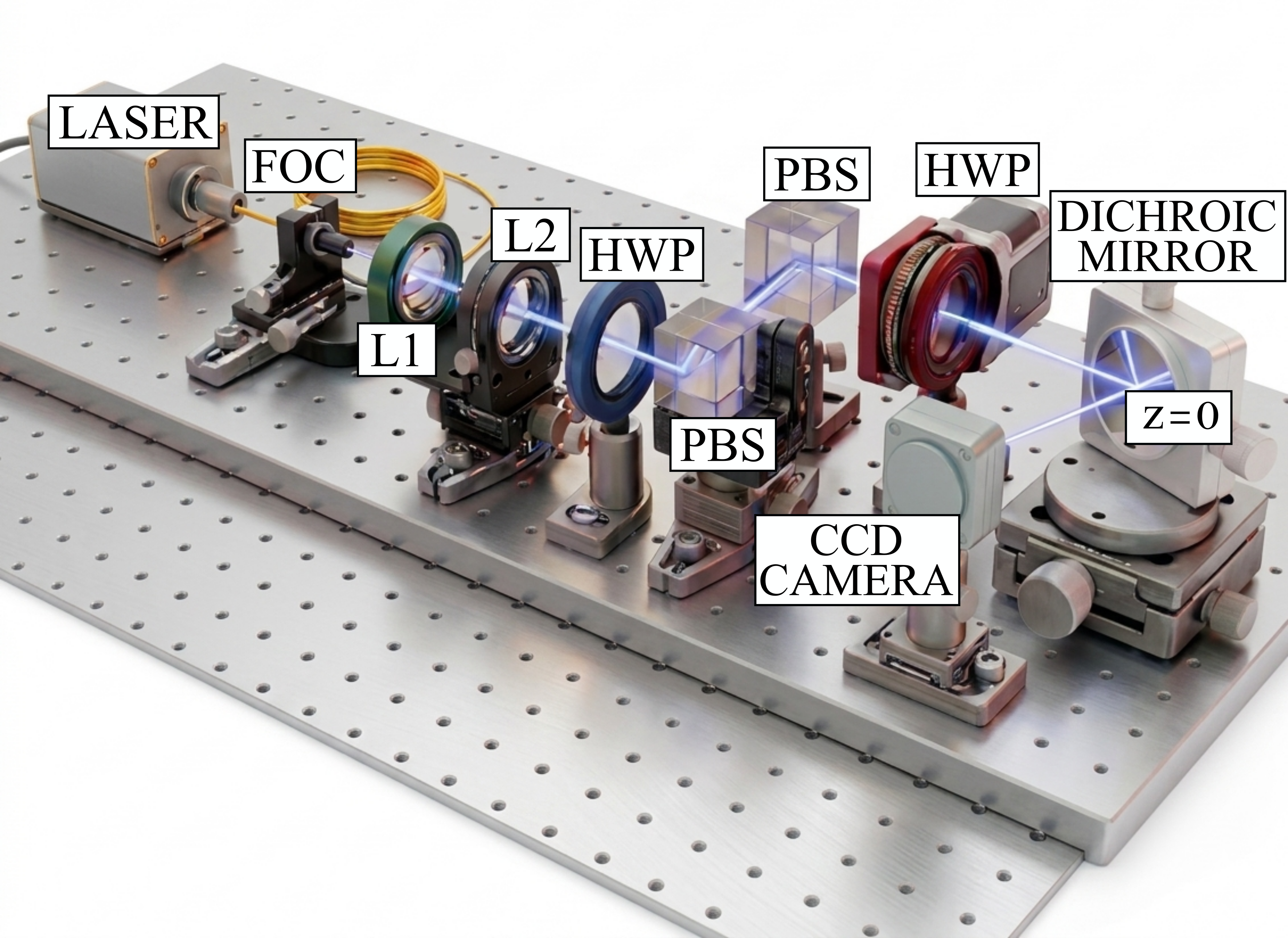}
\caption{(a) Experimental setup (FOC: fibre-optic cable, L1, L2: lenses, HWP: half-wave plate, PBS: polarizing beam splitter).}
\label{fig:setup}
\end{figure}

The experimental setup used to measure the polarization-dependent GH shift is shown in Fig.~\ref{fig:setup}. A continuous-wave laser beam is coupled into a single-mode polarization-maintaining optical fibre to ensure a high-quality fundamental spatial mode. The output beam is collimated and its waist in the mirror position is adjusted using a pair of lenses ($L_{1}$, $L_{2}$), resulting in a beam waist of approximately $w_0 \approx 75~\text{\textmu m}$ and a beam quality factor $M^2 \approx 1$, confirming near-diffraction-limited propagation.

The incident polarization state is controlled using a half-wave plate (HWP) and polarizing beam splitters (PBS). A motorized rotation stage holding an additional HWP enables continuous tuning between the s- and p-polarized incidence on the sample. The dichroic mirror under investigation is mounted on a rotational stage to control the angle of incidence. The reflected beam is imaged onto a CCD camera (pixel size: 5.2~\textmu m). For each angle and polarization, exposure time and attenuation are adjusted to maintain comparable intensities, minimizing systematic centroid bias. The GH shift is determined by a Gaussian fitting of the intensity distribution recorded on the CCD; the fitted centroid corresponds to the beam position at a fixed propagation distance (z\,=\,28\,mm) from the reflection interface. All frames are recorded and processed offline to extract the beam shifts for both polarizations.

The investigated dichroic mirror is designed for deep-ultraviolet (DUV) operation, exhibiting high reflectance in the 220--260~nm spectral range while remaining transmissive outside this band. Both reflection and transmission show strong angular and polarization dependencies, characteristic of multilayer interference coatings. In this work, the fundamental light at 458~\text{nm} is sent to the dichroic mirror, which is used in a frequency-doubling cavity for harmonic separation. At this wavelength, the coating exhibits partial reflection and transmission, enabling phase-sensitive beam shifts in a regime relevant to cavity mode propagation. The detailed optical response at 458~nm is discussed in the following section.

\section{Experimental Results and Discussions}


\subsection{Optical characterization (Reflectance and Transmittance)}

\begin{figure}[t!]
    \centering
    \includegraphics[width=0.8\linewidth]{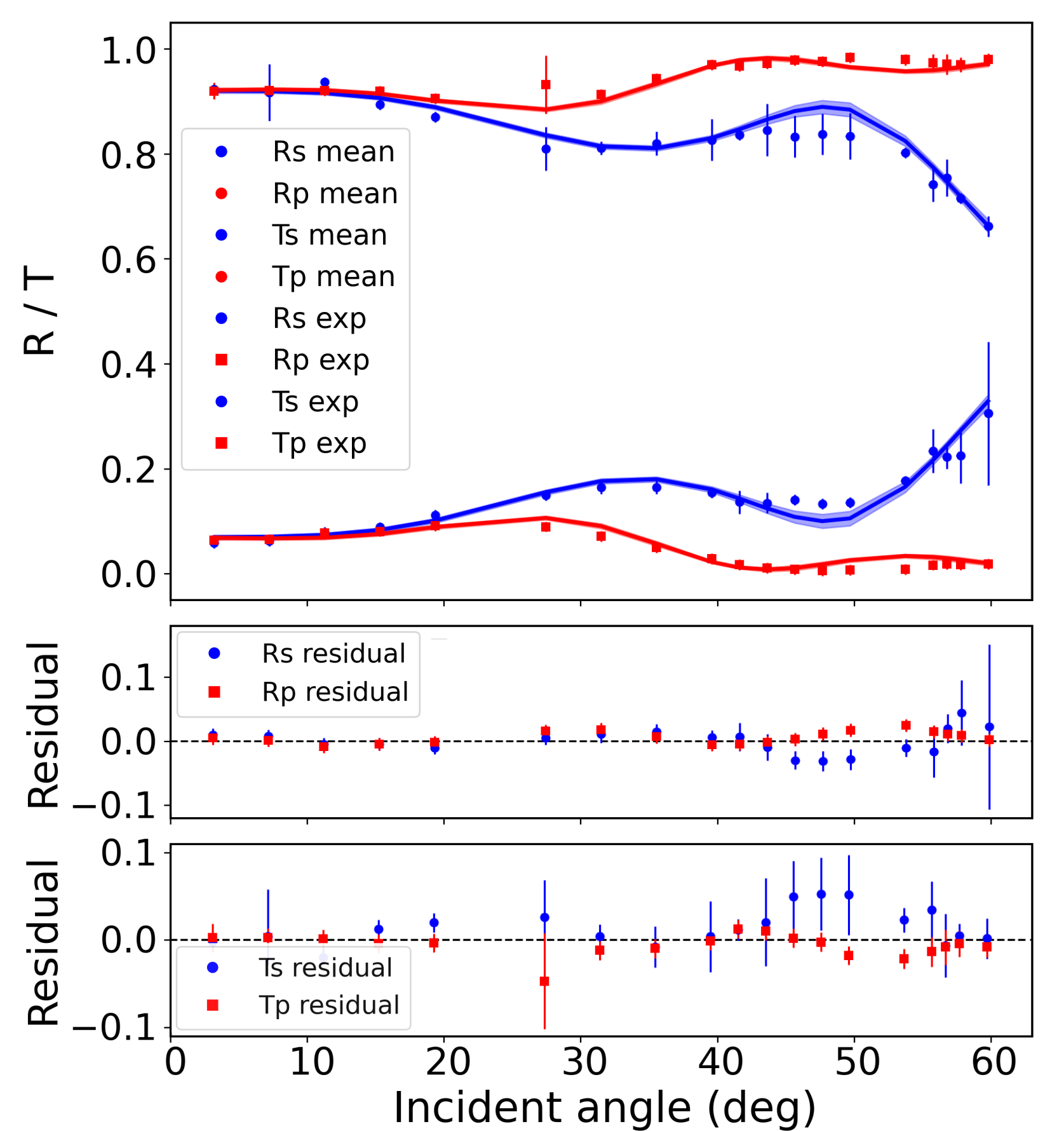}
    \caption{(a) Measured and fitted R and T as a function of incidence angle at 458~nm for s-polarization (blue) and p-polarization (red). Shaded regions indicate uncertainties from bootstrap analysis. The middle and lower panels show residuals for s- and p-polarizations with experimental error bars.}
    \label{fig:RT}
\end{figure}

Angular reflectance and transmittance measurements at 458~nm reveal a strong polarization-dependent characteristic of dielectric coatings. As shown in Figure~\ref{fig:RT}, the p-polarized reflectance exhibits a minimum near the pseudo-Brewster angle region, $\theta_B \approx 46^\circ$, accompanied by enhanced transmittance, consistent with the multilayer nature of the coating, which prevents Rp from vanishing exactly as it would at a single dielectric interface.

In contrast, the s-polarized response remains highly reflective across the full angular range, consistent with constructive interference in multilayer stacks. To quantitatively relate the measured reflectance and transmittance data, a multilayer model based on the transfer matrix method (TMM) was fitted to RT data (see Figure~\ref{fig:RT}). The structure consists of alternating high- and low-refractive-index layers, corresponding to HfO$_2$ and SiO$_2$, respectively, deposited on a substrate. Additional structural details were obtained through communication with the manufacturer; the model presented here serves as an effective representation of the key optical behavior. Layer thicknesses and optical constants were optimized to reproduce the experimental angular response.

Figure~\ref{fig:RT} compares the measured and fitted reflectance and transmittance as a function of the incidence angle for s- and p-polarizations. Shaded regions bootstrap uncertainties of the fitted parameters. The middle and lower panels display the residuals (experiment minus fit) for both polarizations, respectively, with error bars corresponding to the experimental uncertainties.

Importantly, RT measurements constrain only the intensity response and do not uniquely determine the phase of the complex reflection coefficient. Since the GH shift depends directly on the angular derivative of this phase, intensity measurements alone cannot fully characterize the optical response. This motivates direct GH shift measurements.

\subsection{GH shift Measurements in Reflection}

To probe phase-sensitive optical effects beyond intensity-based characterization, the GH shift was measured as a function of the incidence angle for both s- and p-polarized light. The reflected beam was recorded with a CCD camera, and the beam centroid in each frame was extracted by Gaussian fitting. For each polarization state and incidence angle, 20 frames were acquired to improve statistical accuracy. Measurements were repeated while rotating the half-wave plate (HWP) over $360^\circ$ to access all polarization configurations. Before each measurement set, the HWP was calibrated over a full $360^\circ$ rotation to correct systematic offsets and reduce temporal drift. Three independent measurement sets, acquired on different days, were used to assess reproducibility. Statistical uncertainties were estimated from the Gaussian centroid fitting, while beam-pointing stability was evaluated using Allan variance analysis. Uncertainties propagated into the multilayer reconstruction were performed via parametric bootstrap Monte Carlo analysis. Synthetic datasets were generated by adding Gaussian noise consistent with the experimental uncertainties in RT and GH shift data. Each realization was refitted using least-squares optimization, and the resulting ensemble of solutions provides an estimate of the uncertainty in both the reconstructed multilayer structure and the predicted optical response. To isolate the intrinsic GH shift, the differential beam centroid between s- and p-polarized beams was computed with full uncertainty propagation. Systematic HWP offsets were removed, and corrected GH shift values were combined using inverse-variance weighted averaging, with additional variations between polarization configurations included as systematic uncertainties. 

\begin{figure}[t!]
\centering\includegraphics[width=0.8\linewidth]{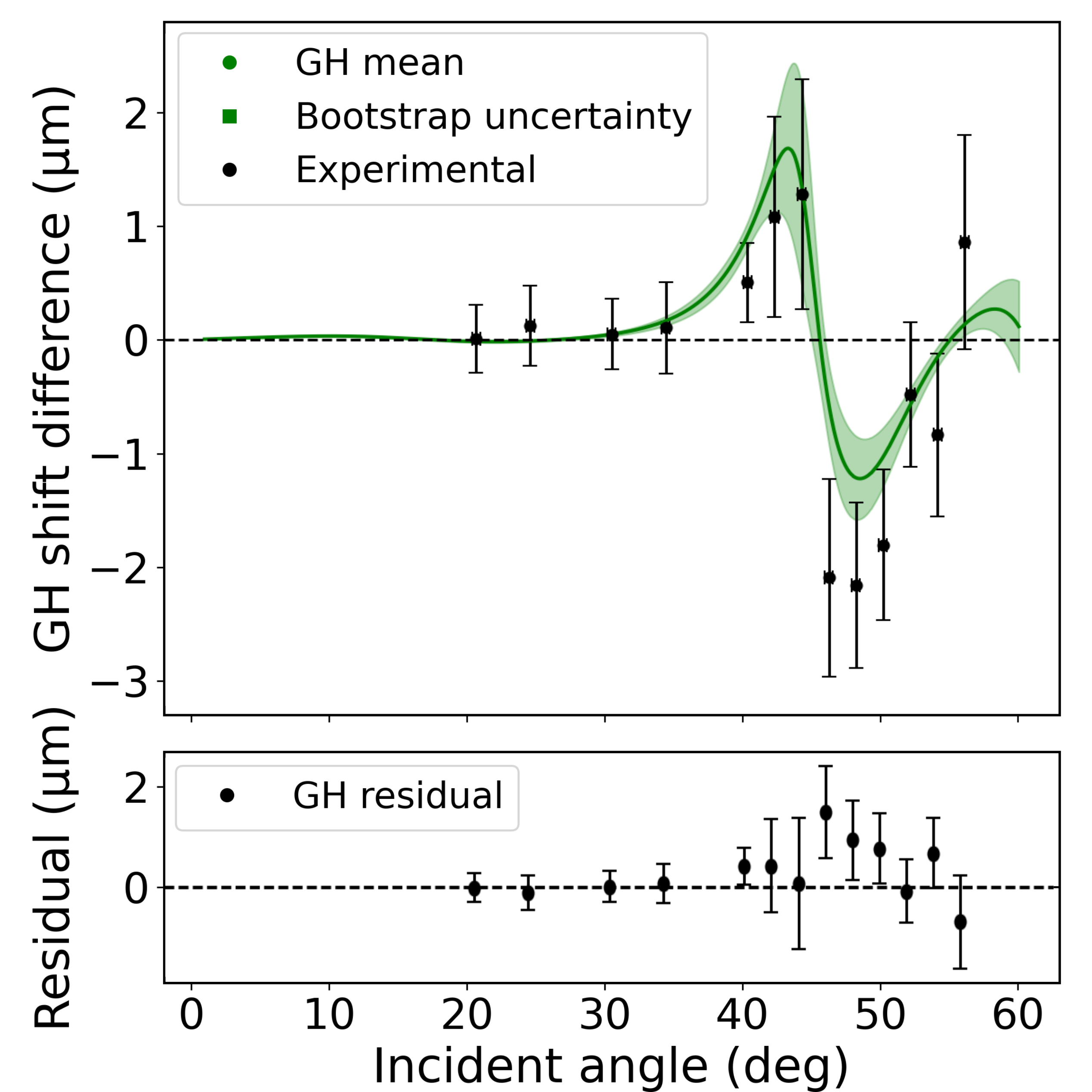}
\caption{Differential GH shift in reflection, $\Delta_{\mathrm{GH}}^{s} - \Delta_{\mathrm{GH}}^{p}$\,, as a function of incidence angle at 458~nm. Error bars denote experimental uncertainties,
solid lines denote the multilayer model.}
\label{fig:ghR}
\end{figure}

Figure~\ref{fig:ghR} shows the resulting differential GH shift ($\Delta_{\mathrm{GH}}^{s} - \Delta_{\mathrm{GH}}^{p}$) at 458~nm as a function of the angle of incidence, measured at a propagation distance of $z~=~28~\mathrm{mm}$ (Rayleigh length $\sim 40~\mathrm{mm}$). Reported uncertainties result from statistical noise, centroid fitting uncertainty, beam drift, and HWP-related systematics, all summed in quadrature. The differential GH shift shows a strong angular dependence, with enhanced magnitude near the angular region where the p-polarized reflectance reaches its minimum. Normalized to the wavelength, the observed peak differential shift of ~2 µm corresponds to $\approx 4.4\lambda$. For comparison, resonantly engineered structures such as magnetoplasmonic photonic crystals have been shown to produce GH shifts up to $\approx18\lambda$ via surface-plasmon excitation~\cite{Makarova2024GHshift}; the comparable order-of-magnitude shift observed here in a non-resonant, commercially relevant DUV dichroic coating underscores that phase-induced beam displacements are not confined to specially engineered resonant systems but are already significant in standard multilayer coatings used in laser cavities.

\begin{figure}[t]
    \centering
    \includegraphics[width=0.75\linewidth]{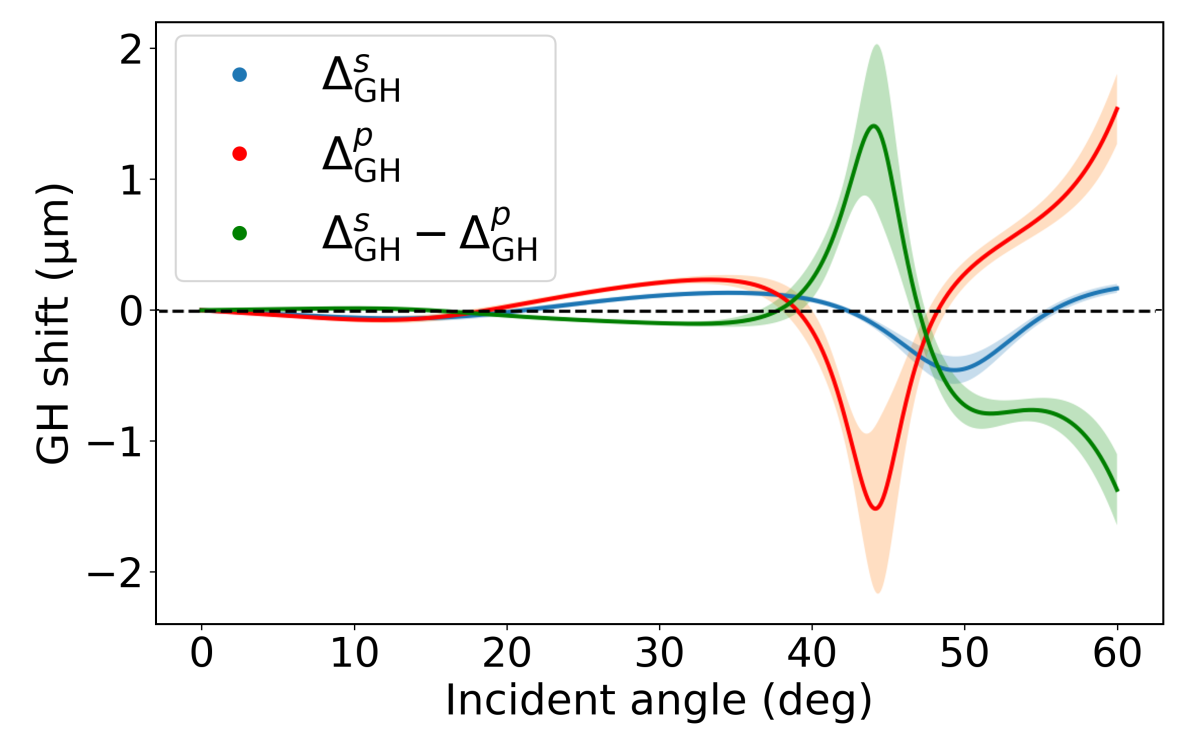}
    \caption{Calculated spatial GH shift for reflected light at the interface ($z~=~0$), as a function of incidence angle.}
    \label{fig:z0}
\end{figure}

A clear sign reversal is observed around this resonance region, with the shift changing from positive to negative before approaching zero at larger angles. The solid curve in Fig.~\ref{fig:ghR} corresponds to the multilayer transfer-matrix model fitted simultaneously to the RT and GH datasets. Good agreement is observed between experiment and model across the full angular range, indicating that the reconstructed multilayer stack captures both amplitude and phase response. Residual analysis shows that most of the data points are consistent with the model within the combined experimental and Monte Carlo uncertainties. Small deviations are observed near the resonance region, where the GH shift is highly sensitive to small variations in the reflection phase and layer thickness.  The observed enhancement and sign reversal of the GH shift can be understood from the phase response of the multilayer coating. Indeed, the GH displacement is proportional to the angular derivative of the reflection phase; thus, rapid phase changes near the resonance region produce large positive or negative beam displacements depending on the sign of the phase gradient. Finally, to separate the intrinsic response and propagation effects, the multilayer model was used to compute the spatial GH shift at the reflection interface ($z~=~0$), shown in Fig.~\ref{fig:z0}. Comparison with the experimentally measured shifts at $z~=~28~\mathrm{mm}$ [See Figure~\ref{fig:setup}] demonstrates that free-space propagation further modifies the observed beam shift through an additional contribution due to propagation.

Once the multilayer structure is constrained by the RT and GH data, the transfer-matrix formalism provides the full complex transmission response, enabling the extraction of polarization-dependent phase shifts through the stack.

\subsection{Lateral Shift in Transmission}

To complement the reflection-based GH analysis described above, the lateral shift in transmission is evaluated using the same transfer-matrix model. In transmission, the observed beam shift is dominated by geometrical propagation through the substrate, while additional contributions arise from the angular-dependent phase of the transmission coefficient.

\begin{figure}[h!]
    \centering
    \includegraphics[width=0.75\linewidth]{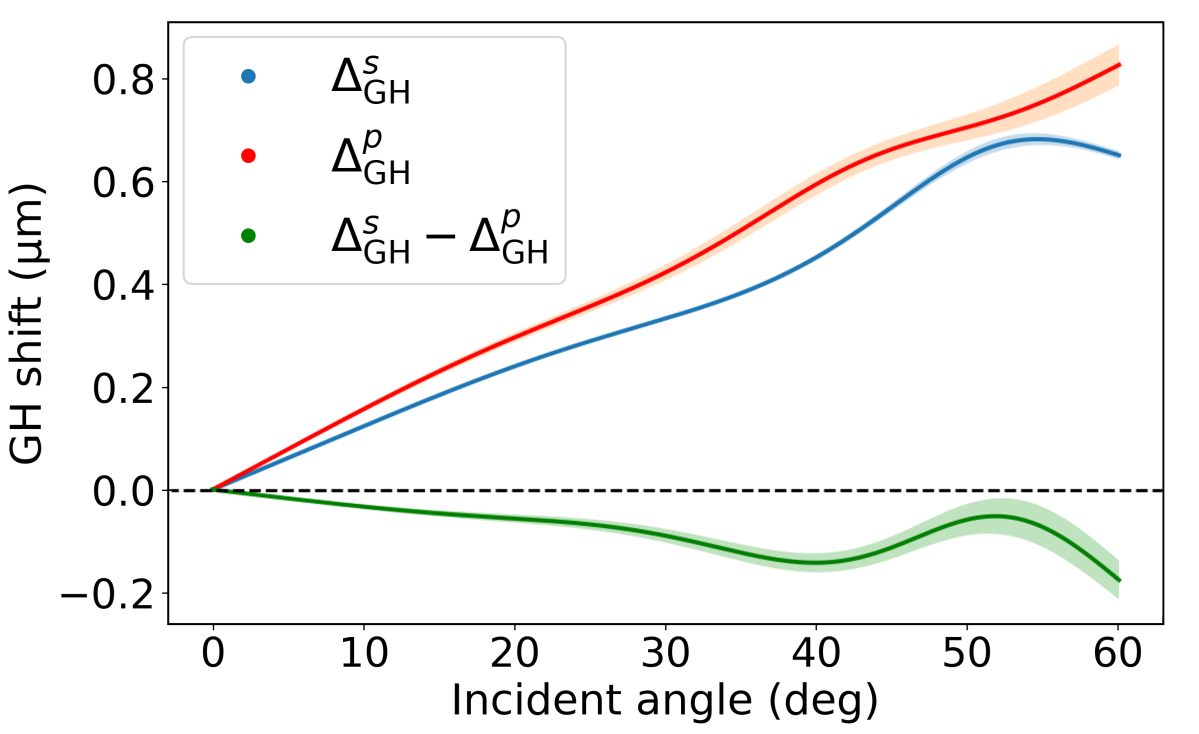}
    \caption{Calculated lateral shift (GH shift) in transmission for s-, p-, and s-p-polarized light as a function of incidence angle.}
    \label{fig:z0t}
\end{figure}

Although the phase-induced contribution is relatively small, it becomes relevant in resonant systems that involve multiple interactions with dichroic mirrors. In optical cavities, such shifts can build up over numerous round trips, modifying the effective propagation axis and intracavity mode overlap. Consequently, the phase response influences cavity dynamics in addition to reflectivity and transmission efficiency. This effect is particularly relevant for frequency doubling cavities for deep-ultraviolet (DUV) production, where operation is highly sensitive to alignment and angular tolerances due to the presence of the phase-matched crystal. While mirror selection is primarily driven by DUV reflectivity, the phase response of the coating offers an additional and often overlooked limitation. This phase behavior is directly manifested in angle-dependent GH shifts, which in high finesse cavities can result in increased sensitivity to alignment and beam pointing. As a result, GH shift characterization provides a complementary metric to evaluate coating performance beyond conventional intensity-based criteria.

\section{Conclusions}

We have demonstrated a combined experimental and modeling approach for characterising a multilayer dichroic mirror, incorporating both intensity- and phase-sensitive optical observables. While angular reflectance and transmittance constrain the amplitude response within a transfer-matrix model, GH shift measurements provide direct access to the phase gradients of complex reflection and transmission coefficients. Their combined analysis provides stronger constraints on the coating's optical response than intensity-only fitting. The results show that dichroic mirrors cannot be treated solely as amplitude elements in cavity modeling. For finite beam waists and non-normal incidence, phase gradients induce measurable beam shifts that modify the propagation axis after successive reflections, impacting intracavity mode geometry, alignment sensitivity, and cavity stability. This work establishes an intensity-phase framework for accurate modeling of optical cavities under realistic conditions. Given the sensitivity of the GH shift to phase response, an interesting direction for future work is to explore its use as a diagnostic for coating aging, contamination, or environmentally induced changes in multilayer structures.

\begin{backmatter}

\bmsection{Funding} PRIN-ISOTOP–2022Z8LX9L, UVQUANT-101080164

\bmsection{Acknowledgment}SM would like to sincerely thank Prof. L. Araujo for his guidance during the early stages of this work and for his constant mentorship, and A. Trenkwalder and the LiCr team for lending the HWP mount.

\bmsection{Disclosures}
Figure 1, illustrating the experimental setup, was initially created manually and refined using AI-assisted image processing (model used is Seedream 5.0 Pro).

\bmsection{Data availability}Data underlying the results presented in this paper and layer parameters used for the fit are not publicly available at this time but may be obtained from the authors upon reasonable request.

\bibliography{sample}

\bibliographyfullrefs{sample}

\end{backmatter}

\end{document}